\documentclass[modern,tighten]{aastex62}

\newcommand{\ns}{{\it NuSTAR}~}
\newcommand{\ir}{{\it IRIS}~}
\newcommand{\sa}{{\it SDO}/AIA~}

\shorttitle{X-ray, EUV and UV Observations of a Small Microflare.}

\shortauthors{Hannah et al.}

\begin{document}

\title{\textbf{\large Joint X-ray, EUV and UV Observations of a Small Microflare}}

\author[0000-0003-1193-8603]{Iain G. Hannah}
\affiliation{SUPA School of Physics \& Astronomy, University of Glasgow, Glasgow G12 8QQ, UK}

\author{Lucia Kleint}
\affiliation{University of Applied Sciences and Arts Northwestern Switzerland, 5210 Windisch, Switzerland}
\affiliation{Leibniz-Institut f\"ur Sonnenphysik (KIS), Sch\"oneckstrasse 6, 79104 Freiburg, Germany}

\author{S{\"a}m Krucker}
\affiliation{University of Applied Sciences and Arts Northwestern Switzerland, 5210 Windisch, Switzerland}
\affiliation{Space Sciences Laboratory University of California, Berkeley, CA 94720, USA}

\author{Brian W. Grefenstette} 
\affiliation{Cahill Center for Astrophysics, 1216 E. California Blvd, California Institute of Technology, Pasadena, CA 91125, USA}

\author{Lindsay Glesener}
\affiliation{School of Physics \& Astronomy, University of Minnesota - Twin Cities, Minneapolis, MN, 55455, USA}

\author{Hugh S. Hudson}
\affiliation{SUPA School of Physics \& Astronomy, University of Glasgow, Glasgow G12 8QQ, UK}
\affiliation{Space Sciences Laboratory University of California, Berkeley, CA 94720, USA}

\author{Stephen M. White}
\affiliation{Air Force Research Laboratory, Space Vehicles Directorate, Kirtland AFB, NM 87123, USA}

\author{David M. Smith}
\affiliation{Santa Cruz Institute of Particle Physics and Department of Physics, University of California, Santa Cruz, CA 95064, USA}

\correspondingauthor{I. G. Hannah}
\email{iain.hannah@glasgow.ac.uk}

\begin{abstract}
We present the first joint observation of a small microflare in X-rays with the Nuclear Spectroscopic Telescope ARray ({\it NuSTAR}), UV with the Interface Region Imaging Spectrograph ({\it IRIS}) and EUV with the Solar Dynamics Observatory/Atmospheric Imaging Assembly ({\it SDO}/AIA). These combined observations allow us to study the hot coronal and cooler chromospheric/transition region emission from the microflare. This small microflare peaks from 2016 Jul 26 23:35 to 23:36UT, in both {\it NuSTAR}, {\it SDO}/AIA and {\it IRIS}. Spatially this corresponds to a small loop visible in the {\it SDO}/AIA \ion{Fe}{18} emission, which matches a similar structure lower in the solar atmosphere seen by {\it IRIS} in SJI1330\AA~and 1400\AA. The {\it NuSTAR} emission in both 2.5-4 keV and 4-6 keV, is located in a source at this loop location. The {\it IRIS} slit was over the microflaring loop, and fits show little change in \ion{Mg}{2} but do show intensity increases, slight width enhancements and redshifts in \ion{Si}{4} and \ion{O}{4}, indicating that this microflare had most significance in and above the upper chromosphere. The {\it NuSTAR} microflare spectrum is well fitted by a thermal component of 5.1MK and $6.2\times10^{44}$ cm$^{-3}$, which corresponds to a thermal energy of $1.5\times10^{26}$ erg, making it considerably smaller than previously studied active region microflares. No non-thermal emission was detected but this could be due to the limited effective exposure time of the observation. This observation shows that even ordinary features seen in UV can remarkably have a higher energy component that is clear in X-rays.
 \end{abstract}

\keywords{Sun: X-rays, gamma rays --- Sun: UV radiation --- Sun: activity --- Sun: corona}

\section{Introduction}

Microflares are small releases of stored magnetic energy in the solar atmosphere that heat material and accelerate particles. Energetically, they are down to a million times smaller than the largest events, yet still demonstrate similar properties \citep{2011SSRv..159..263H}. The smaller microflares range down to GOES A-Class events, with 1-8\AA~ flux $<10^{-7}$ Wm$^{-2}$, and are considerably more frequent than the largest flares. The frequency distribution of flares is a negative power-law with index $\alpha \sim 2$ \citep{1991SoPh..133..357H}. However, it is still not clear down to what energy scales this rate persists, a crucial fact to determine the overall contribution of micro-, or even smaller, nanoflares, to heating the solar corona.

X-ray observations of microflares provide clear diagnostics of the energetics of the heated material and accelerated electrons. Above a few keV this is predominantly bremsstrahlung continuum emission and {\it RHESSI} \citep{2002SoPh..210....3L} showed that microflares down to the GOES A1 level exhibit non-thermal footpoints, at the ends of coronal loops containing material $>10$MK \citep{2002SoPh..210..445K,2008A&A...481L..45H}. A large statistical study of {\it RHESSI} microflares \citep{2008ApJ...677.1385C,2008ApJ...677..704H} showed that these {\it RHESSI} events were exclusively in active regions, lasted for a few minutes, were not necessarily spatially small, had emission $>10$MK, and over the initial impulsive period had median thermal energy $10^{28}$ erg and non-thermal $10^{27}$ erg. 

Going beyond {\it RHESSI}, the Nuclear Spectroscopic Telescope ARray \ns \citep{2013ApJ...770..103H} is an astrophysics telescope with two direct focusing optics modules, and a higher effective area than {\it RHESSI}. It has targeted the Sun several times since the first pointing in late 2014 \citep{2016ApJ...826...20G}, and has observed active region microflares as well as quiet Sun brightenings. These microflares are about an order of magnitude weaker than {\it RHESSI} could observe, down to an estimated GOES level of $\sim$A0.1, and showed heating up to about 10MK, with thermal energies of $10^{27}$erg \citep{2017ApJ...844..132W,2017ApJ...845..122G}. These events were also well observed at longer wavelengths, in softer X-rays with {\it Hinode}/XRT and the Solar Dynamics Observatory/Atmospheric Imaging Assembly (\sa). X-ray brightenings were also observed with \ns outside of active regions in the quiet Sun, with temperatures of about 3-4MK, estimated GOES emission of $\lesssim$A0.01 and thermal energy of about $10^{26}$ erg \citep{2018ApJ...856L..32K}.

These microflares and brightenings did not show any non-thermal emission, but this is likely due to an observational constraint of using \ns to observe the Sun. \ns has a detector throughput of only 400 counts s$^{-1}$ telescope$^{-1}$, which even low levels of solar activity can swamp \citep{2016ApJ...826...20G}. This results in a detector livetime fraction considerably less than unity, and a greatly reduced effective exposure time. Given the steep nature of a typical solar X-ray spectrum, these short effective exposure times limit the spectral dynamic range, producing few, or no, counts at higher energies, the range in which non-thermal emission is expected \citep{2016ApJ...820L..14H}.  The effective exposures of these \ns observations were short because there were also other bright sources on the solar disk. Unfortunately even regions outside the \ns field of view of 12' $\times$ 12' can be detected and exacerbate the throughput issue \citep{2016ApJ...826...20G}. Therefore the best observing conditions with \ns are during periods of low overall solar activity, with the brightest feature within the \ns field of view. Even with higher livetime \ns observations there is still a limit to the sensitivity arising from the inherent short duration of these small flares.

Non-thermal emission is expected from small microflares as particle acceleration often features during magnetic reconnection, the energy release mechanism that is thought to be behind flares of all sizes. Even the smallest {\it RHESSI} microflares could show considerable non-thermal emission from accelerated electrons \citep{2008A&A...481L..45H}. Electrons accelerated in small impulsive events are thought to be behind coronal radio emissions such as Type I noise storms \citep{1997ApJ...474L..65M,2011SoPh..273..309S}, however it is considerably more difficult to obtain the electron energetics from the coherent radio emission, compared to X-rays, due to the non-linear nature of the emission mechanism processes. The presence of accelerated electrons in small events has also been inferred from UV observations with \ir \citep{2014SoPh..289.2733D}. Rapid brightenings (over 10s of seconds) were observed at the footpoints of hot coronal loops \citep{2014Sci...346B.315T}. The observed blue-shifts (upflows) of the \ion{Si}{4} 1403\AA~line in these ``moss'' brightenings \citep{1999SoPh..190..409B} are consistent with RADYN numerical simulations of chromosphere/transition region heating by a beam of accelerated electrons (a power-law of non-thermal energy $6\times10^{24}$ erg, with spectral index $\delta=7$ above a cutoff of $E_{C}=10$keV). Thermal conduction and Alfven waves could not reproduce the line shift, nor the intense brightening \citep{2014Sci...346B.315T}. This combination of RADYN simulations and UV observations were further shown to provide constraints to the properties of the non-thermal electrons \citep{2018ApJ...856..178P}. They found that the blue-shifts were dependent on both the non-thermal energy and the low energy cutoff otherwise red-shifts were produced; $E_{C}\ge5$keV for $10^{24}$ erg, and $E_{C}\ge15$keV for $10^{25}$ erg. This work showed that \ion{Mg}{2} could also be used to help constrain the electron beam properties. 

In this paper, we present observations of a small microflare that occurred on 2016 Jul 26 at 23:35 (SOL2016-07-26T23:35) in X-rays with \ns, UV with \ir and EUV with \sa, allowing us to study the heating of both the chromosphere/transition region and corona. In \S\ref{sec:over}, we give an overview of the event, before going into detail about the spatial and temporal behaviour of the microflare in \S\ref{sec:imgtimp}. Then in \S\ref{sec:spec} we derive properties of the emission from both the \ir and \ns spectra. The thermal properties found from the \ns spectra are compared to the emission observed by \sa and {\it GOES}/XRS

\begin{figure}
\centering
\includegraphics[width=10cm]{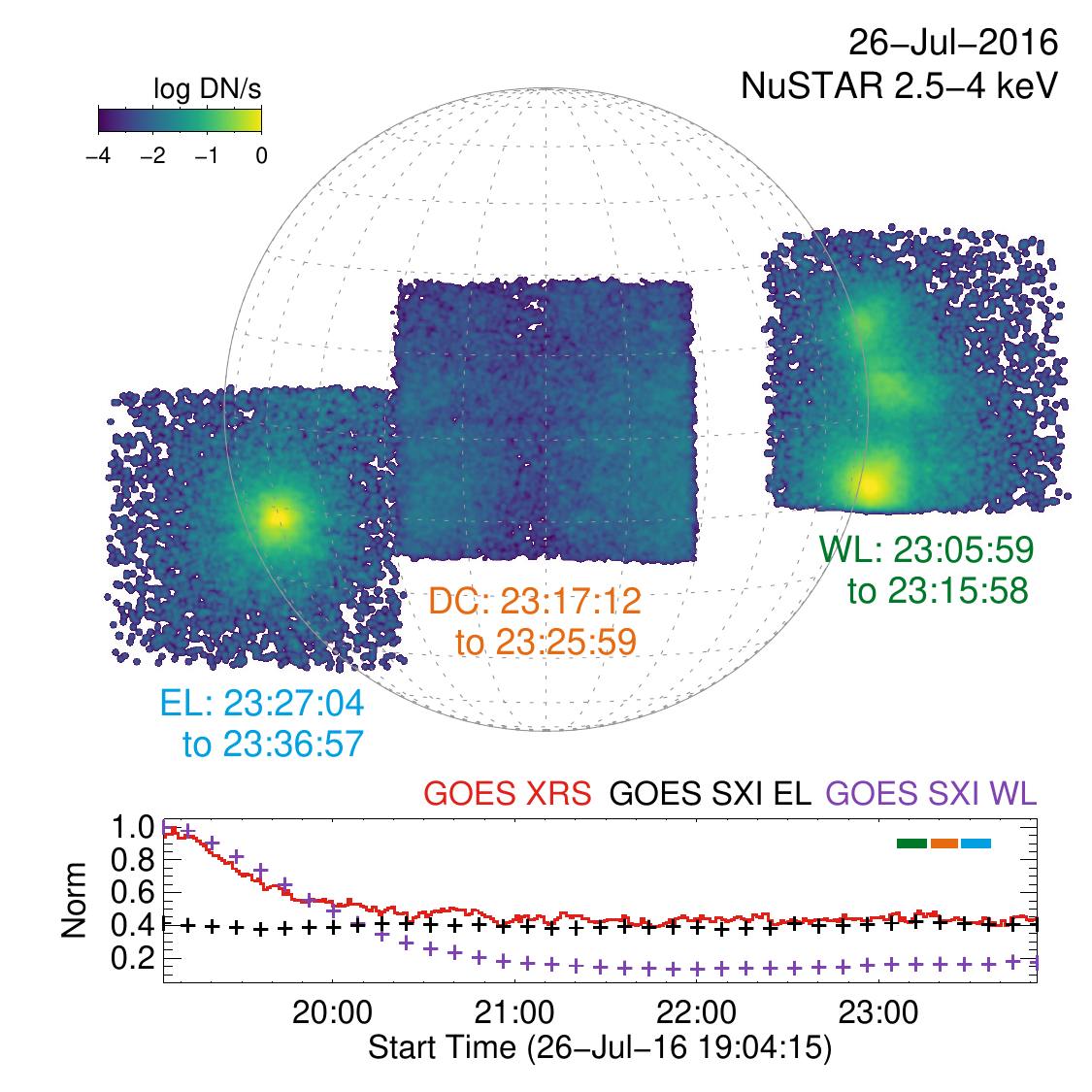}
\caption{(Top) Full solar disk view of the \ns pointing on 2016 Jul 26, showing 10 minutes of data from the occulted regions at the west limb (WL) pointing, before the disk centre (DC) target, and then the region of interest in this paper on the east limb (EL). (Bottom) Time profile of the Soft X-rays from the full-disk via {\it GOES}/XRS 1-8\AA~(red line) and from just the WL (purple crosses) and EL (black crosses) regions using {\it GOES}/SXI ``Be12a'' wavelength.}
\label{fig:img_fd}
\end{figure}

\section{Observation Overview}\label{sec:over}

\begin{figure}
\centering
\includegraphics[width=8cm]{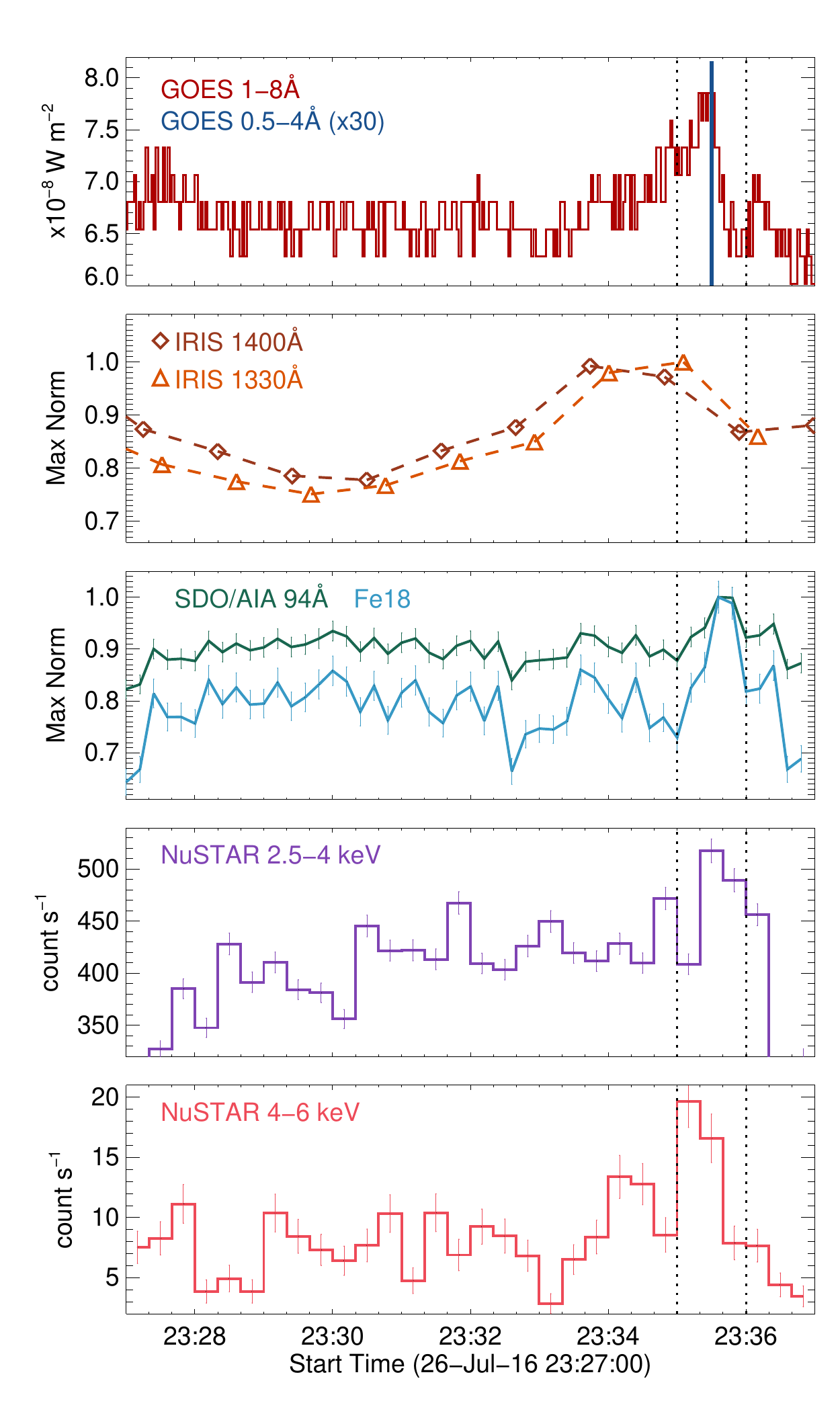}
\caption{The microflare time profile from (top to bottom) {\it GOES}/XRS 1-8\AA~and 0.5-4\AA, \ir SJI1400\AA~and SJI1330\AA, \sa \ion{Fe}{18}, \ns 2.5-4 keV and 4-6 keV. The dotted vertical lines indicate the time of the microflare over 23:35 to 23:36UT. Here the GOES lightcurves (top panel) is from the full-disk emission, whereas the other panels are the box regions highlighted in Figure \ref{fig:imgs} for \sa and \ir and in Figure \ref{fig:imgs_ovr} for \ns.}
\label{fig:ltc}
\end{figure}

\begin{figure}
\centering
\includegraphics[width=15cm]{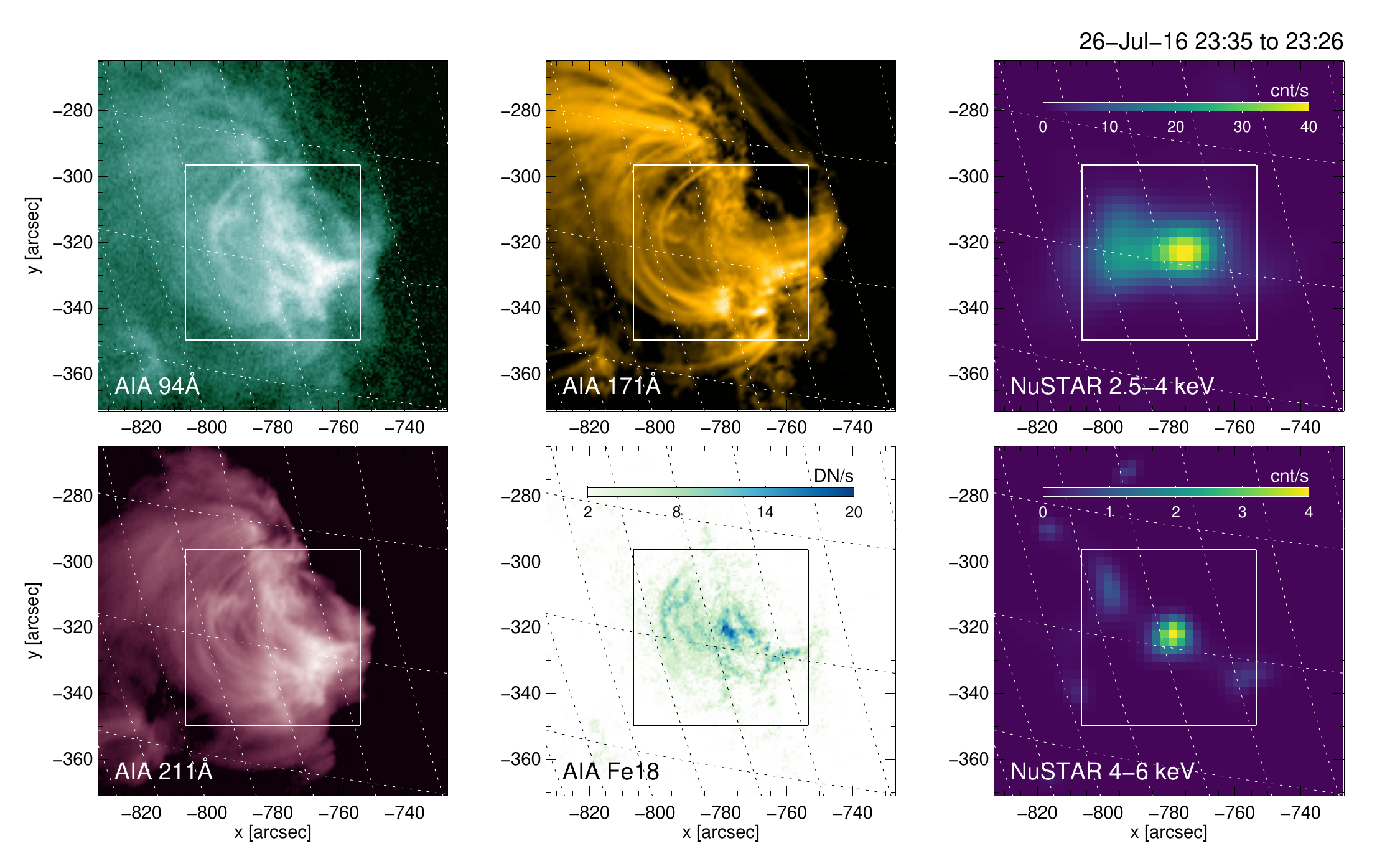}
\caption{An overview of the microflaring region seen in \sa 94\AA, 171\AA, 211\AA, \ion{Fe}{18}, and \ns 2.5-4 and 4-6 keV. Shown is the emission summed over the microflare time 23:35 to 23:36UT. The box indicates the part of the region with the hottest emission, over which the \ns lightcurve and spectrum are produced, see Figures~\ref{fig:ltc} and \ref{fig:ns_spect}.}
\label{fig:imgs_ovr}
\end{figure}

The region that produced the microflare was observed by \ns on 2016-Jul-26 between 23:27 and 23:37UT, towards the south-eastern limb. The region was never given an NOAA ID, but was identified as SPoCA 19717 \citep{2014A&A...561A..29V}. This particular \ns solar pointing\footnote{For an overview of \ns solar pointings see \url{http://ianan.github.io/nsovr}} had spent 3 hours focused on active regions on the opposite western limb as they rotated off the visible disk, occulting the brighter emission from the lower solar atmosphere. \ns briefly pointed at disk centre, during which it observed a small quiet Sun event reported in \citet{2018ApJ...856L..32K}, before targeting the eastern region for 10 minutes. It then returned to the west limb for another hour. The resulting \ns images are shown in Figure~\ref{fig:img_fd}, as well a time profile of the full-disk soft X-ray emission seen by {\it GOES}/XRS 1-8\AA~and from just the occulted west limb (WL) and east limb (EL) using {\it GOES}/SXI ``Be12a'' wavelength. By the time \ns is observing the eastern region it is the brightest X-ray source on the disk, with the other regions being well-occulted. Therefore, what is seen in the {\it GOES}/XRS full-disk emission should be dominated by the target region. Both the 1-8\AA~and 0.5-4\AA~channels of {\it GOES}/XRS show a small microflare between 23:35 to 23:36UT; see the top panel of Figure~\ref{fig:ltc}. This event peaks at {\it GOES} A8-level, but is only about an A1 excess above the pre-flare emission. The full time profile and spatial behaviour of the microflare in X-ray, EUV, and UV are shown in Figures~\ref{fig:ltc}, \ref{fig:imgs_ovr} and \ref{fig:imgs} will be discussed in the next section, \S\ref{sec:imgtimp}. {\em Hinode}/EIS was also targeting this region, but at the time of the \ns observations the slit was further to the east of the microflare, missing the hot material \ns was detecting.

\section{Spatial \& Temporal Analysis}\label{sec:imgtimp}

\begin{figure*}
\centering
\includegraphics[width=15cm]{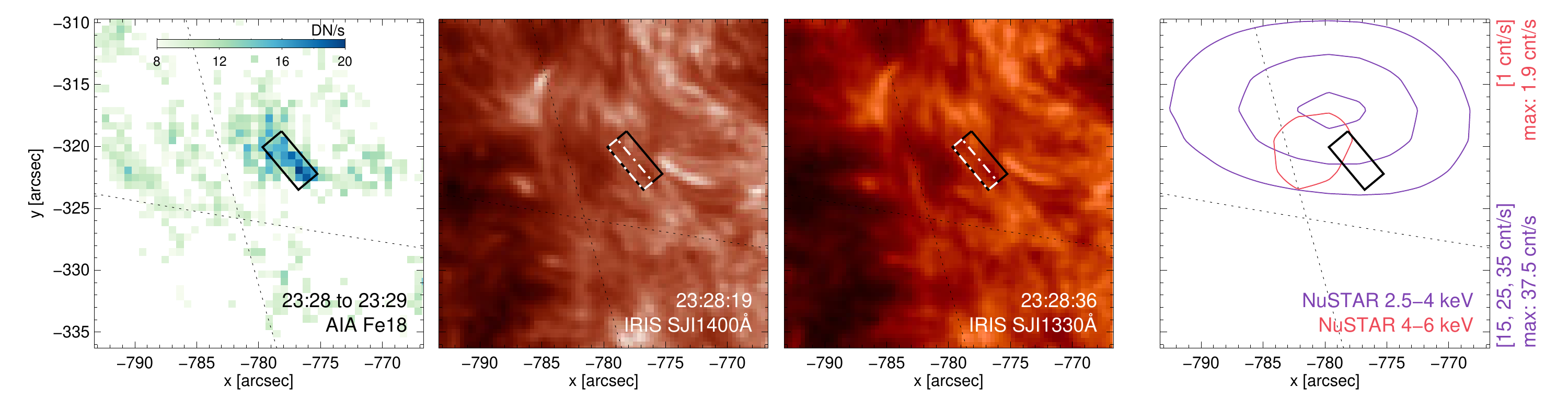}\\
\includegraphics[width=15cm]{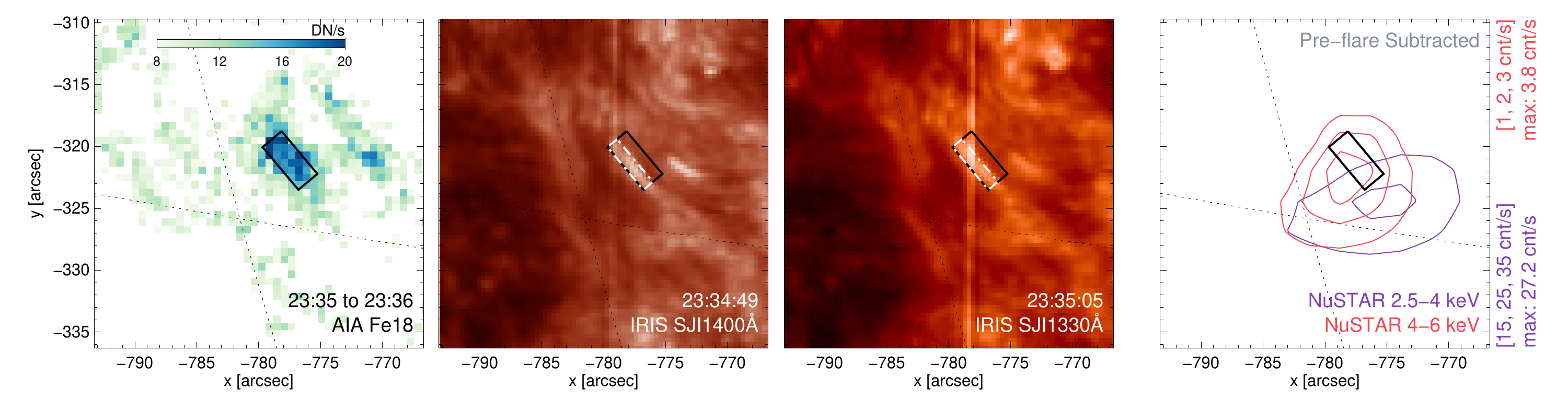}
\caption{The microflaring region seen in \sa \ion{Fe}{18}, \ir SJI1400\AA, SJI1330\AA~and contours of the \ns 2.5-4 and 4-6 keV emission (from left to right). Shown are the pre-flare time over 23:28 to 23:29UT (top row) and the time of the microflare, 23:35 to 23:36UT (bottom row). The solid black rectangle in all panels highlights the location of the \sa \ion{Fe}{18} loop, the dashed-dotted white rectangle in the middle panels indicates the smaller \ir loop, both used to produce the lightcurves in Figure~\ref{fig:ltc}. The same absolute contour levels are used for both times in each \ns energy band, given to the right of the panel with the max value of each image. Note that the \ns contours for the microflaring time (bottom right panel) has the pre-flare image subtracted.}
\label{fig:imgs}
\end{figure*}

The \sa images of the region were processed to level 1.6 data, using the standard software to prep, as well as deconvolving the point spread function. Most of the EUV channels showed only weak emission from the region of interest, which barely changed over the 10 minutes. The 94\AA~channel did show a small loop and this brightened at the same time as the X-ray emission. We removed the cooler component of the  94\AA~channel via the approach of \citet{2013A&A...558A..73D}, leaving just the emission above 3~MK from \ion{Fe}{18} . Overview images of the whole region in \sa 94\AA, 171\AA~and 211\AA, as well as \ion{Fe}{18} are shown in Figure~\ref{fig:imgs_ovr} for the microflare time, 23:35 to 23:36UT. The region has extensive cooler emission, but the hotter emission, as indicated by \ion{Fe}{18}, is more compact with the small ``loop'' that brightens during the microflare time. A zoomed-in view of this emission, shown in the left-hand panels of Figure~\ref{fig:imgs}, gives both the pre-flare (23:28 to 23:29UT) and microflare (23:35 to 23:36UT) times, where the loop and the brightening becomes clearer. In both Figures~\ref{fig:imgs_ovr} and \ref{fig:imgs}, the \sa images shown here have been summed over 1 minute to improve the signal-to-noise ratio. The full 12s cadence images were used to determine the time profile of the EUV emission from just the brightening loop region. The resulting lightcurves (shown in the second top panel of Figure~\ref{fig:ltc}) clearly show a peak of emission, and just slightly after the X-ray microflare seen with {\it GOES}/XRS. Figure~\ref{fig:imgs} does show that there is some increase in \ion{Fe}{18} outside of the loop considered however this emission is not as bright as the loop nor does it have the same time profile as the X-ray microflare. The 304\AA~channel had some features that brightened over the observation time but were not spatially or temporally correlated with the 94\AA~nor \ion{Fe}{18} emission.

IRIS co-observed the region from 2016-Jul-26 21:53:26 to 2016-Jul-27 02:47:17UT with 17 large, sparse, 64-step rasters (OBSID 3600110059) with steps of 1\arcsec, exposure times of 15s, and a factor 2 for spatial and spectral summing. All SJI filters (1330\AA, 1400\AA, 2796\AA, 2832\AA) were used, giving a cadence of 65 s for each SJI filter. For this paper, we analyze the 6th raster, taken from 23:19:57 to 23:36:59UT. We verified the remaining orbital variation to be below 0.2 km s$^{-1}$ in near UV during this raster, which is below our desired accuracy, and therefore we use the original raster with the newest calibrations (L12-2017-04-23) for the analysis without additional corrections. We align the SJI1400\AA~data to the \sa 1600\AA~data, which includes a 0.6$^\circ$ roll and a $<$2\arcsec\ shift. There are multiple little bright loops in the \ir SJI1400\AA~ and SJI1330\AA~images, including one at exactly the same location, and of the same length and orientation of the microflaring loop seen in \sa \ion{Fe}{18}. This is shown in the middle panels of Figure~\ref{fig:imgs}. The lightcurves from this small UV loop (shown in Figure~\ref{fig:ltc}) indicate that it brightens but does so slightly before the time of the X-ray and EUV microflare. Crucially the \ir slit moves across the loop during the time of the microflare, from 23:35 to 23:36UT, and this spectral analysis is detailed in \S\ref{sec:spec}.

 The \ns emission from the region is taken only from the FPMA telescope, as the FPMB telescope has the detector gap directly through the microflare. The \ns data were filtered\footnote{For software to work with the \ns solar data see \url{https://github.com/ianan/nustar_sac}} to remove bad pixels, and non-grade 0 events to minimise detector pileup \citep{2016ApJ...826...20G}. These observations were all made with pointing determined by a single Camera Head Unit (CHU3), so we can apply just one correction to the data. The \ns pointing was aligned to the whole region seen in \sa \ion{Fe}{18}, not just the microflaring loop. The resulting lightcurves for the \ns emission in 2.5-4 keV and 4-6 keV were found over this larger region (shown in Figure~\ref{fig:imgs_ovr}) and both energy ranges brighten during the microflare time.  The emission peaks slightly earlier in the higher energy channel, consistent with flare heating followed by cooling. The \ns images for both energy channels, and the pre-flare and microflare times, are shown as maps in Figure~\ref{fig:imgs_ovr} and zoomed in contours in Figure~\ref{fig:imgs}. These images have been processed to deconvolve the point spread function however it will be only partially removed, so the \ns images and contours are likely still larger than the true source size.
 
 The \ns images during the microflare (see Figure~\ref{fig:imgs_ovr}) show a bright source with centroid close to the \ion{Fe}{18} loop in both 2.5-4 keV and 4-6 keV. The 2.5-4 keV image does show a more extended structure which approximately matches the larger structure of fainter loops seen in \ion{Fe}{18}, to the east of the microflaring loop. To isolate the microflare source in the \ns images from the ``background'' active region emission we subtracted the pre-flare images from the microflare images, shown in Figure~\ref{fig:imgs}. The pre-flare \ns contours (top right panel Figure~\ref{fig:imgs}) show an extended 2.5-4 keV source, with centroid slightly to the north of the microflaring loop, and a weak compact 4-6 keV source near the loop. The \ns microflare contours with the pre-flare subtracted (bottom right panel Figure~\ref{fig:imgs}) show a brighter compact source in 4-6 keV overlapping the microflaring loop location. This is unsurprisingly similar to the non-subtracted image (shown in Figure~\ref{fig:imgs_ovr}) as the pre-flare source in this energy range is very weak. The microflare is dominating the emission in the 4-6 keV energy range - something we further confirm via the \ns spectral fitting in \S \ref{sec:spec}. There is a more substantial change in the pre-flare subtracted 2.5-4 keV image; the source of the excess emission due to the microflare is more compact with centroid very close to the  loop position. This strengthens the argument that the X-ray microflare is coming from a source smaller than seen by \ns and is highly likely to be the loop structure seen in EUV and UV. Although the \ns 2.5-4 keV and 4-6 keV centroids do not perfectly align with each other, or the loop,  this is not a significant difference as it is within the spatial resolution of \ns - the full width at half-maximum of the optics' point spread function is 18''.

\section{Spectral Analysis}\label{sec:spec}
\subsection{IRIS spectra}

\begin{figure}
\centering
\includegraphics[width=14cm]{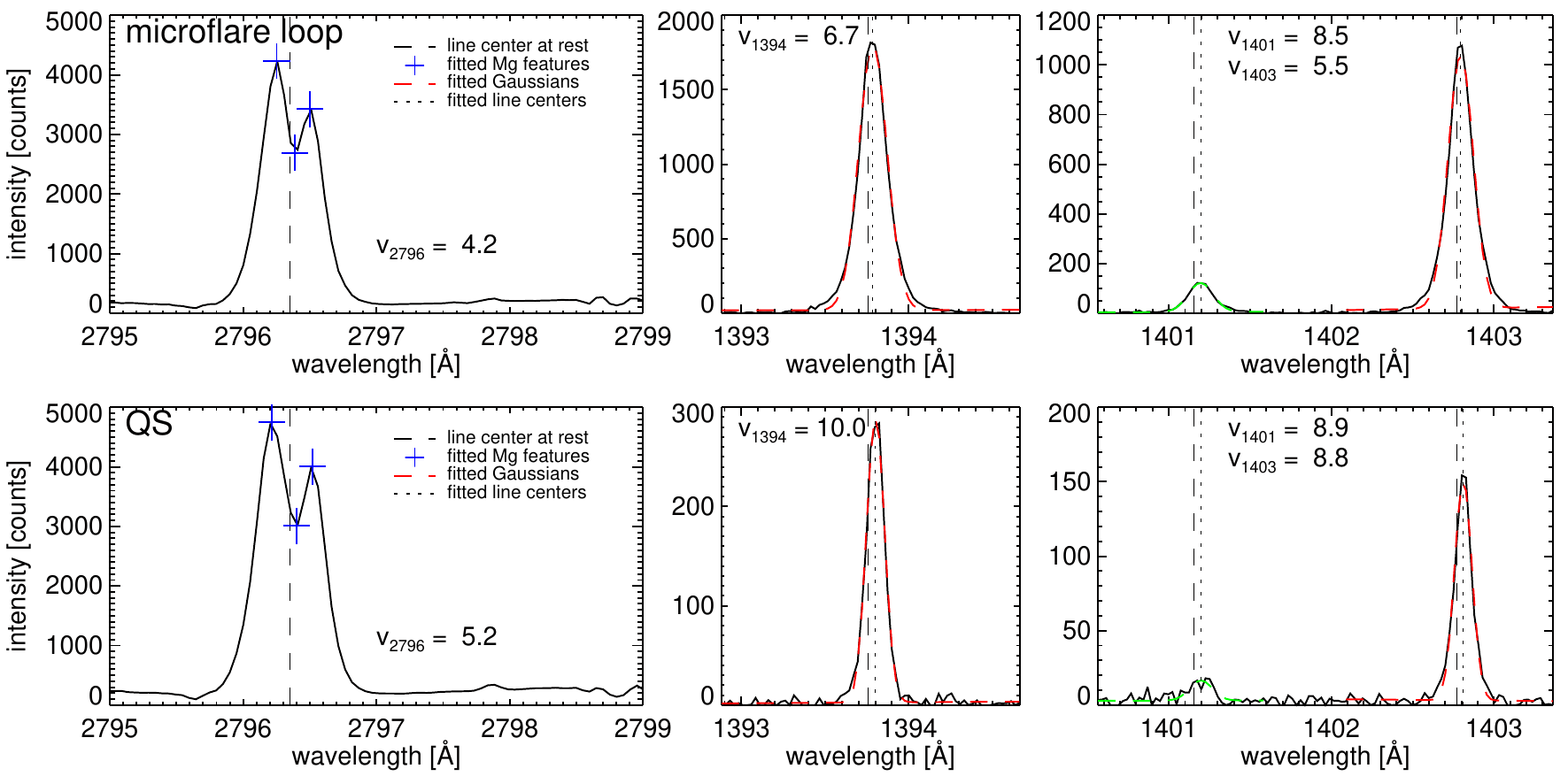}
\caption{Examples of fits to \ir~line profiles. The top row shows a pixel in the microflare, the bottom row a quiet Sun pixel. The Doppler velocities are given in km s$^{-1}$. For Mg (left column), the feature locations and intensities (blue crosses) are found automatically, for the other spectral lines, regular Gaussian fits are performed (dashed coloured lines).}
\label{examplefits}
\end{figure}

At the time of the \ns observation the \ir slit was over the microflaring loop and moved across it during the 1 minute it brightened in X-rays and EUV, but faded in UV. To obtain the features of the spectral lines we use the \textit{iris\_get\_mg\_features\_lev2.pro} routine for Mg, which derives the position and intensities of the blue, red and central peaks. We mainly focus on Mg $k$ at 2796.35\AA, because the $h$ line shows identical behaviour. For the FUV lines, we perform Gaussian fitting to the \ion{Si}{4} 1393.76\AA, \ion{Si}{4} 1402.77\AA, and \ion{O}{4} 1401.16\AA\ lines, and obtain the Doppler shifts, Doppler widths, and line intensities. Examples of such fits are shown in Figure~\ref{examplefits} for a pixel in the microflare (top row) and a quiet Sun pixel (bottom row). In these example fits the microflare is nearly an order of magnitude brighter in \ion{Si}{4} and \ion{O}{4} than the quiet Sun pixel shown, but the \ion{Mg}{2} intensity is similar. The velocities for this pixel of the microflare show that they are not significant compared to the chosen quiet Sun pixel. The other \ion{O}{4} lines (1399.78\AA, 1404.78\AA) are below the noise limit in most pixels and therefore cannot be used as a density diagnostic. The coronal \ion{Fe}{21} line, which usually only appears in flares, is below the detection limit in this microflare.

\begin{figure*}
\centering
\includegraphics[width=14cm]{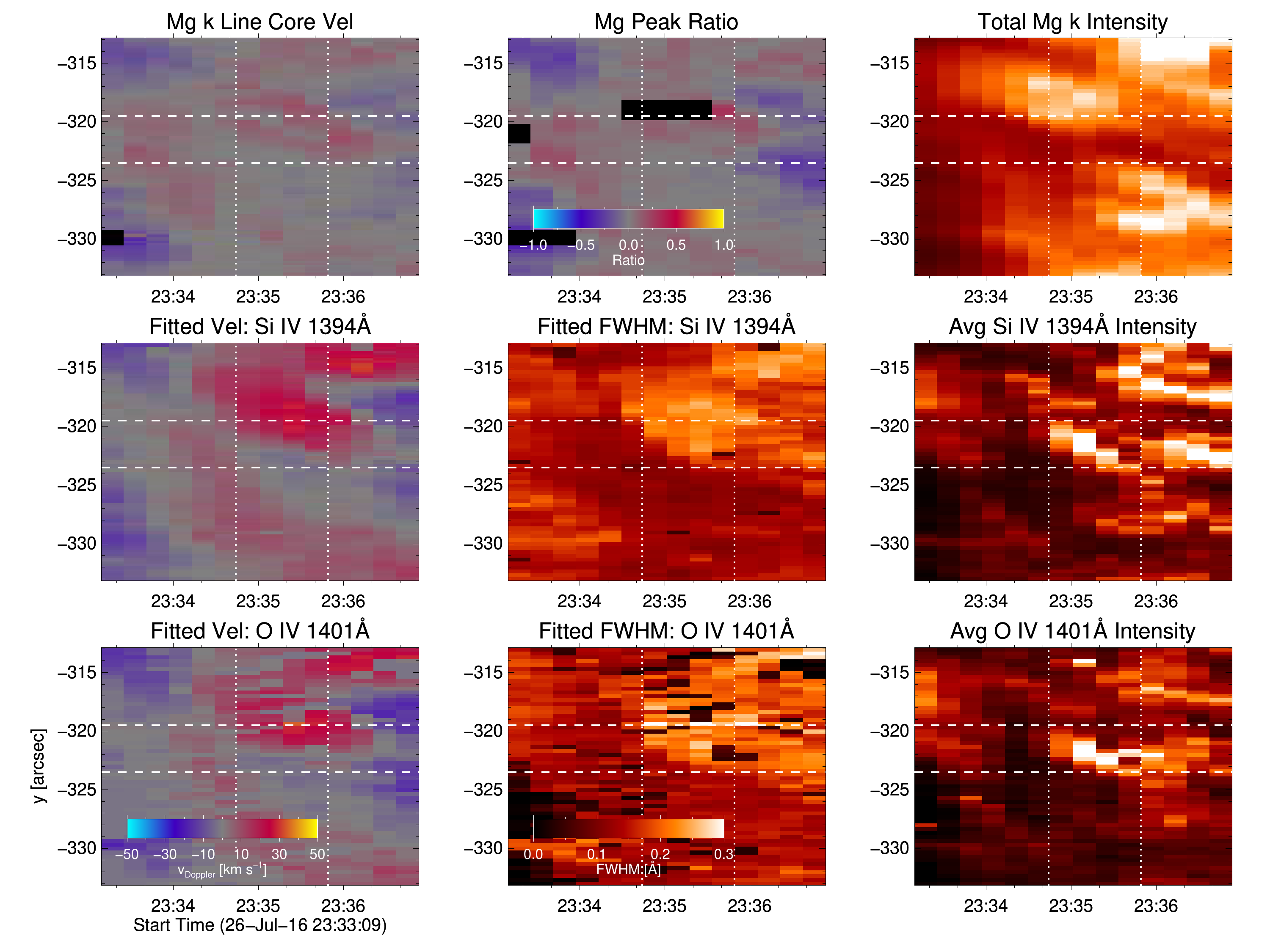}
\caption{Spectral fit results to the IRIS slit spectrum, using the fitting approach as shown in Figure~\ref{examplefits}. The top row show the line core velocity, peak ratio and total intensity for \ion{Mg}{2}. The middle and bottom rows show the fitted velocity, line FWHM and intensity for \ion{Si}{4} and \ion{O}{4} respectively. The dotted vertical lines show the time range the slit is over the microflaring loop. The dashed horizontal lines show the vertical extent of the loop.}
\label{fig:iris_spec}
\end{figure*}

Maps of the fit results are shown in Figure~\ref{fig:iris_spec}. The box formed by the white lines indicates the location and time interval of the microflare. The \ion{Mg}{2} line core Doppler velocity around the microflare shows a weak redshift of less than 10 km s$^{-1}$, which occurs everywhere in the field of view. The Mg peak ratio is defined in Eq.~2 of \citet{2013ApJ...772...90L} and it correlates with the average velocity in the upper chromosphere. Black areas indicate fitting issues, i.e. locations where the Mg line profiles do not show their typical shape, but rather a single peak. The peak ratio around the microflare is zero indicating that it does not influence the apparent upper chromospheric dynamics. Similarly, it is invisible in the Mg intensity. \ion{Si}{4} 1394\AA~and 1403\AA~are very similar, therefore only \ion{Si}{4} 1394\AA~is shown in the plots, as it has a higher absolute intensity. Note that the ratio of the total intensity in these two \ion{Si}{4} lines for the microflare loop are approximately 2, the expected value for optically thin emission \citep{2018arXiv181111075K}. In \ion{Si}{4} the microflare is clearly visible in the intensity maps. The Doppler width of \ion{Si}{4} is slightly enhanced (0.2\AA), but such enhancements also occur in other parts of the FOV and can therefore not be attributed solely to the microflare. The velocities of \ion{Si}{4} are generally higher than those of \ion{Mg}{2}, as can be expected, because \ion{Si}{4} forms at higher temperatures. At the location of the microflare downflows of the order of 20 km s$^{-1}$ are prevalent, which are commonly found in the quiet Sun. The microflare is also visible in the \ion{O}{4} intensity maps. The \ion{O}{4} 1401\AA~line is often weak, which explains the lack of fits (black locations) in its FWHM plot. Similarly to \ion{Si}{4}, the \ion{O}{4} FWHM and velocities are enhanced, but it is unclear if this is related to the microflare because similar enhancements are seen throughout the quiet Sun. The fact that the small loop is visible in the \ion{O}{4} and \ion{Si}{4} lines suggests material that is heated to log$T=4.8$ and log$T=5.2$ (or 0.06 and 0.16MK) respectively. Because the loop is invisible in the \ion{Mg}{2} line, it means that there is little material at log$T=4.0$ in the loop. It seems that plasma below the upper chromosphere is not significantly affected by this microflare.

\begin{figure}
\centering
\includegraphics[width=7cm]{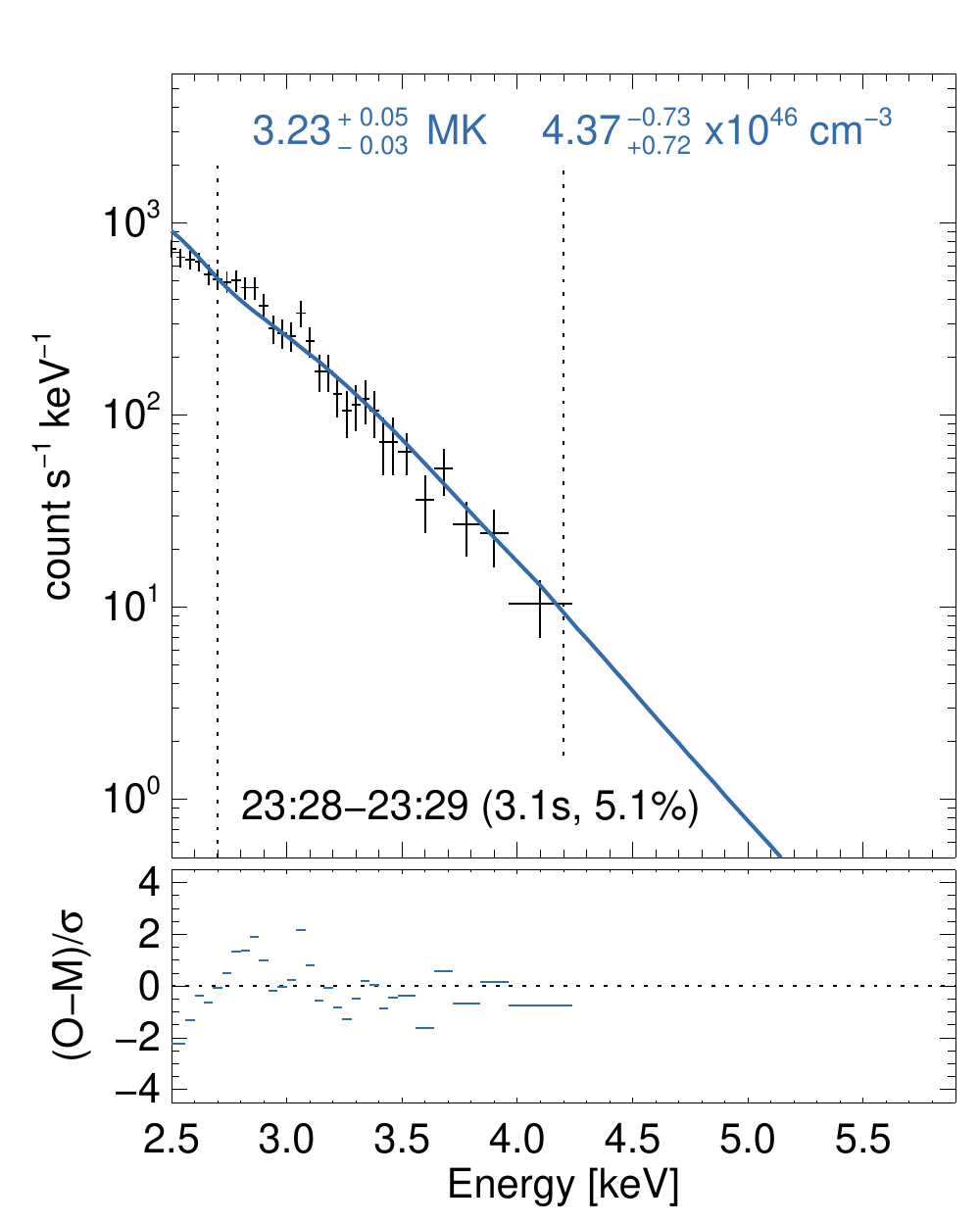}
\includegraphics[width=7cm]{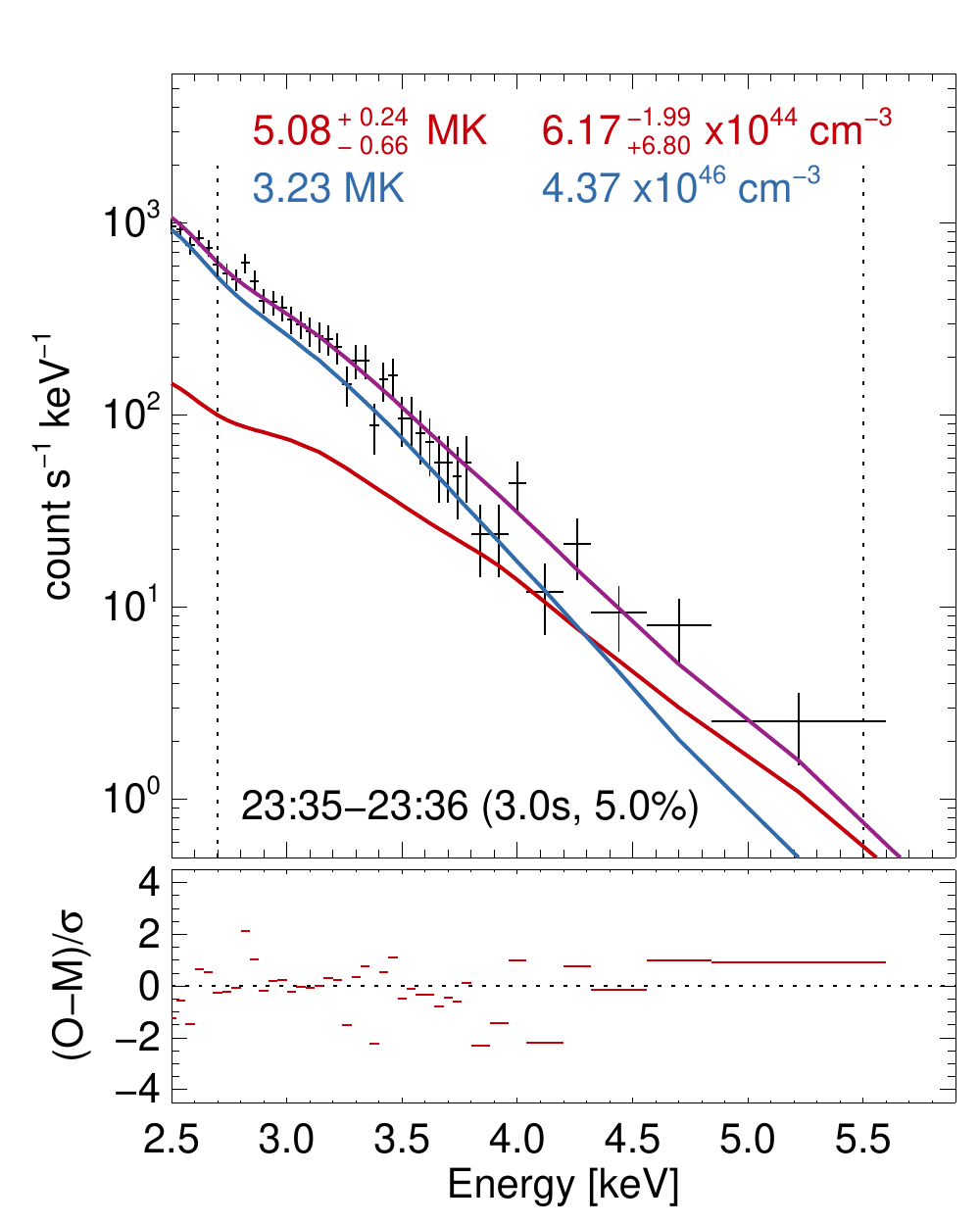}
\caption{\ns spectra (black crosses) and fits for the pre-flare (left) and microflare (right) times. The pre-flare spectrum (left) is fitted with a single thermal APEC model (blue line), with parameters of temperature and emission measure. The microflare spectrum (right) is fitted with two thermal APEC model components (purple line), one using fixed values found from the pre-flare time (blue line), the other fitting the excess (red line). The vertical dotted lines indicate the energy range each fit was performed over. The bottom panels shows the residuals of the fits. Note that the error is not symmetric, with the largest temperature corresponding to the smallest emission measure.}
\label{fig:ns_spect}
\end{figure}

\subsection{\ns spectrum}
The fitted \ns spectrum for the pre-flare and microflare times are shown in Figure~\ref{fig:ns_spect}. Here we show the spectrum from \ns FPMA over the region shown in Figure~\ref{fig:imgs_ovr} during 23:28 to 23:29UT and 23:35 to 23:36UT. The data was rebinned before fitting so that there were at least 10 counts in each bin. Bad pixels and non-zero grade events were filtered out of the eventlist used to make the spectrum. The spectra were fitted in XSPEC using the APEC thermal model, with coronal abundances manually set using the values from \citet{1992ApJS...81..387F}, not using the default solar ones (which are photospheric and not coronal). The minimum fit energy used was 2.7~keV, as below this energy there is a discrepancy in the instrumental response arising from uncertainty in the detection threshold \citep{2018SPIE10709E..2VG}. The best fit parameters were found using the Cash statistic \citep{1979ApJ...228..939C}. The pre-flare spectrum (left panels Figure~\ref{fig:ns_spect}) is well fitted with a single thermal component of temperature 3.23MK and emission measure $4.37\times10^{46}$ cm$^{-3}$. These fit parameters are coupled and not symmetric, so the minimum temperature within the error range corresponds to the largest emission measure: i.e. $1\sigma$ uncertainty of 3.20MK corresponds to $5.09\times10^{46}$ cm$^{-3}$ and 3.28MK corresponds to $3.64\times10^{46}$ cm$^{-3}$. We use this pre-flare spectral fit to take account of the emission from the rest of the region (background) during the microflare time. To fit the microflare excess above the pre-flare emission we added a second thermal component and found a good fit to the data with an additional component of 5.08MK and $6.17\times10^{44}$ cm$^{-3}$ (right panels Figure~\ref{fig:ns_spect}). The $1\sigma$ uncertainty ranges for these parameters are 4.41MK and $1.30\times10^{45}$ cm$^{-3}$ and 5.32MK and $4.18\times10^{44}$ cm$^{-3}$. Taking the uncertainty in the pre-flare fit into account does not significantly change the fit obtained for the excess during the microflare. There are no solar counts above 5.5keV in this event, from hotter or non-thermal emission, but this observation did have a short effective exposure (about 3.0s from an ontime of 60s and livetime about 5.0\%) and the microflare was only well observed in one of the two telescopes, limiting the spectral dynamic range.

Using the observed \sa \ion{Fe}{18} loop, of about 8 pixels long by 4 pixels wide, we get a volume estimate of $8.3\times10^{24}$ cm$^{3}$, assuming a filling factor of unity. This is smaller than the \ns observed source size, shown in Figure~\ref{fig:imgs}. However, as we discussed in \S\ref{sec:imgtimp}, the \ns images are likely larger than the true emitting region due to the 18'' full width at half-maximum of the optics' point spread function, with the deconvolution approach only partially reducing this blurring effect. We therefore assume that the \ns source size matches the smaller \sa \ion{Fe}{18} loop as it is more representative of the true source size, an approach that has been used several times before \citep[c.f.][]{2017ApJ...844..132W,2017ApJ...845..122G,2018ApJ...856L..32K}. This volume combined with the \ns emission measure gives a density of $8.64\times10^9$ cm$^{-3}$, with an uncertainty range of $7.11\times10^9$ cm$^{-3}$ to $1.25\times10^{10}$ cm$^{-3}$.

From this we can calculate the instantaneous thermal energy \citep{2008ApJ...677..704H} of the microflare over the minute it is seen above the pre-flare emission, finding  $1.50\times10^{26}$ erg, with an uncertainty range of $1.30\times10^{26}$ erg to $1.90\times10^{26}$ erg. This means that this event is about an order of magnitude smaller in energy than active region microflares previously seen with \ns \citep{2017ApJ...844..132W,2017ApJ...845..122G}. \ns observations of quiet Sun flares \citep{2018ApJ...856L..32K} showed a similar thermal energy to the microflare presented in this paper. The density and thermal energy of the whole region during the pre-flare time can also be estimated by assuming the volume of the region is related to the observed \sa \ion{Fe}{18} area as $V=A^{3/2}$, giving a density of $8.73\times10^8$ cm$^{-3}$ and thermal energy of $1.05\times10^{29}$ erg. So the microflaring loop contains only about 0.14\% the thermal energy of the whole region and is not contributing substantially to the overall heating of the region.

\begin{figure}
\centering
\includegraphics[width=7cm]{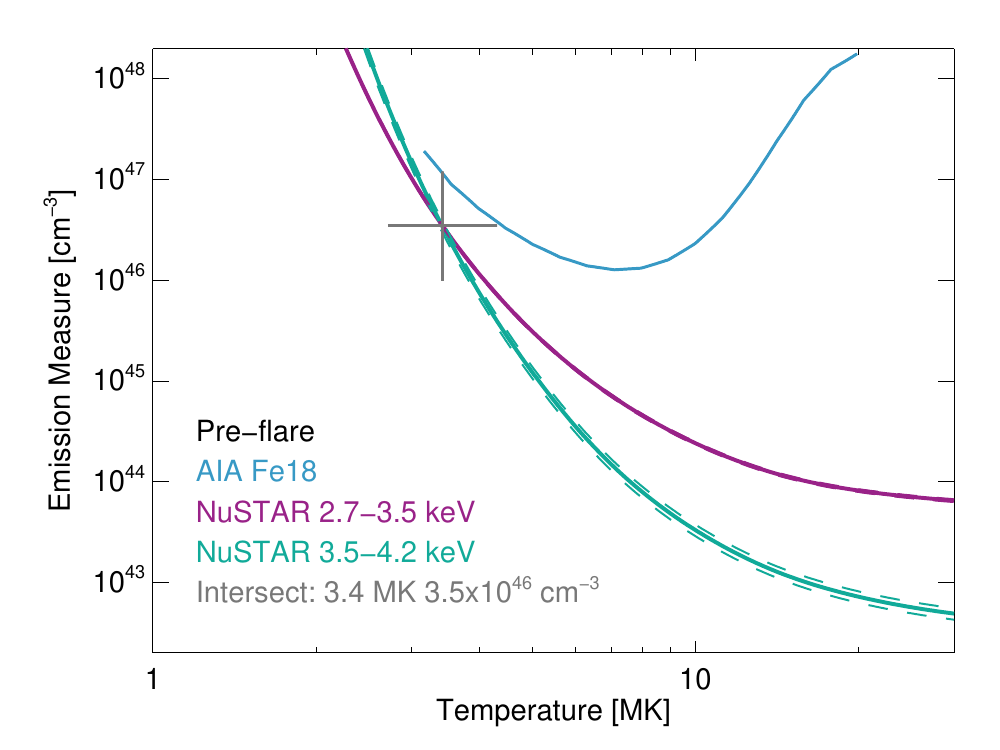}
\includegraphics[width=7cm]{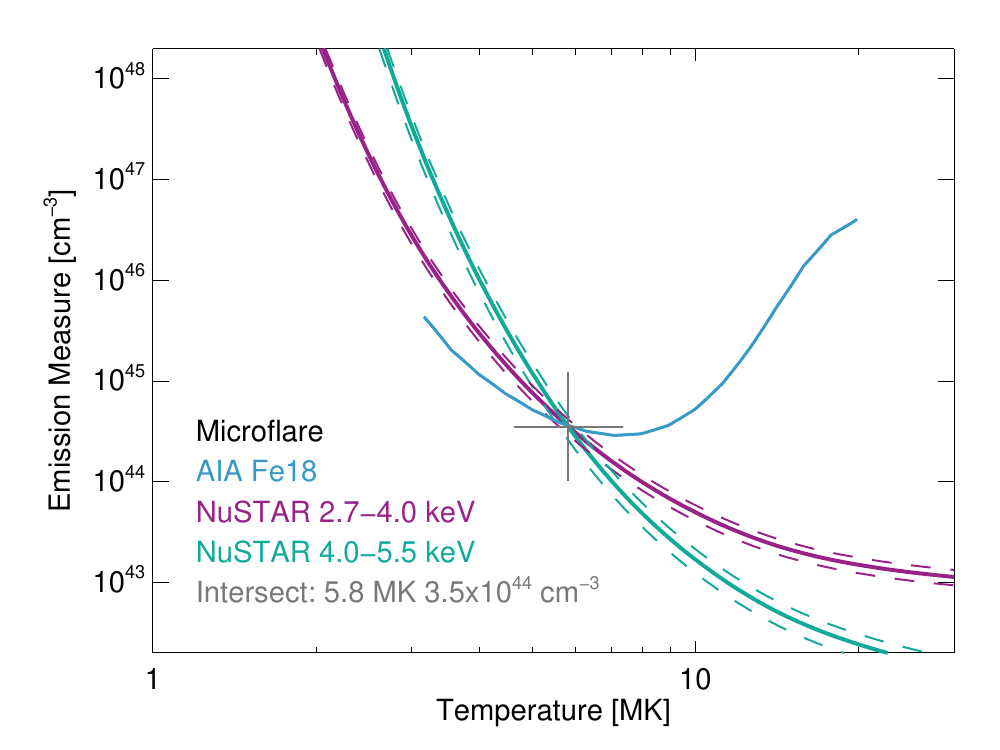}\\
\caption{EM loci curves for the pre-flare (left) and microflare excess (right). Shown are the loci curves for two \ns energy ranges, and \sa \ion{Fe}{18}. The dashed lines give the bounds of the uncertainty for each curve. In both panels, the grey cross shows the intersect of the two \ns energy bands, the temperature and emission measure of this point given in the legend.}
\label{fig:emloci}
\end{figure}

\subsection{Comparison of \ns and \sa}

By calculating the \ns count rate and the thermal response
%\footnote{\url{https://github.com/ianan/nustar_sac/blob/master/idl/make_nstresp.pro}} 
in two different \ns energy ranges, we can produce the EM loci curves (the rate divided by the response). These determine the maximum possible emission measure for each isothermal temperature and can help verify the thermal parameters found from the \ns spectral fitting. They can also be used to show whether the emission observed by \ns and \sa are coming from the same thermal source. The resulting EM loci curves are shown in Figure~\ref{fig:emloci} for the pre-flare and microflare times. Different energy ranges are used for each time interval, determined from approximately the mid-point of the fit range of the spectra (Figure~\ref{fig:ns_spect}). For the pre-flare time 2.7 to 3.5~keV and 3.5 to 4.2~keV is used and these two EM loci curves intersect at a slightly higher temperature and lower emission measure than was found from spectral fitting. This consistency between the EM loci and spectral fitted values is despite the APEC thermal model being using for the fitting, and CHIANTI atomic database for the EM loci curves. There is a mismatch between the \sa and \ns curves but that is likely due to the \ns observed emission being at the edge of \ion{Fe}{18} temperature response range. Also the calculation of the \ion{Fe}{18} emission is an empirical approach and does not perform well when the emission is weak, such as we have in this region. For the microflare time, we want to determine the thermal parameters of the excess over the pre-flare time, so subtract the earlier emission. The resulting EM loci curves for both the \ns and \sa \ion{Fe}{18} channel all intersect at the same temperature and emission measure, showing that both instruments are observing the same loop material at around 5-6MK. Again the temperature from the EM loci approach is higher than the spectral fitting, and the emission measure is lower, but still consistent.

\begin{figure}
\centering
\includegraphics[width=8cm]{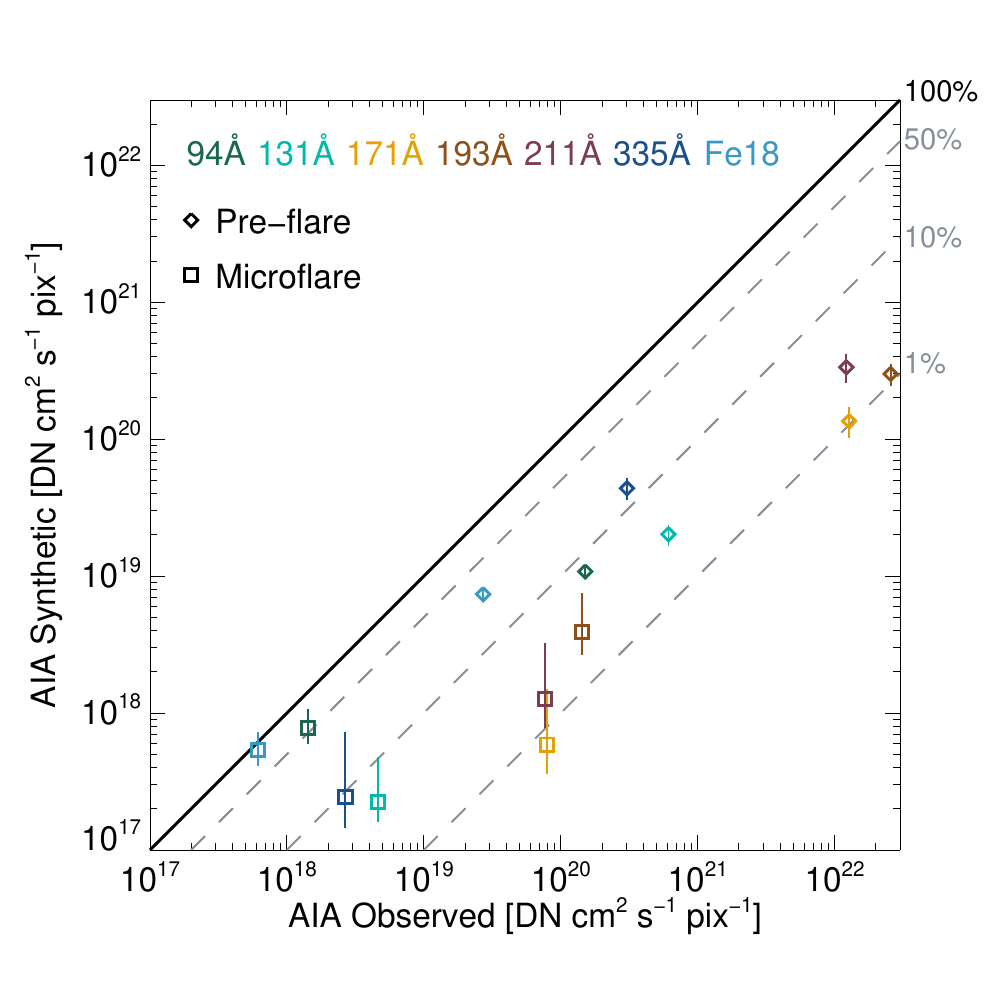}
\caption{The observed \sa emission versus the synthetic fluxes derived from the \ns spectral fit (AIA synthetic). The different colours indicate the \sa channel, and the different symbols the pre-flare and microflare time. The solid black line shows where 100\% of the observed \sa would be produced by the material observed by \ns. The grey dashed lines show where the \ns observed emission would be contributing  50\%, 10\% or 1\% of the emission observed with \sa.}
\label{fig:synobs}
\end{figure}

For a clearer comparison of the observed \ns and \sa emission, we take the temperature and emission found from fitting the \ns spectrum and fold this through the temperature response for each \sa channel. We then compare the observed \sa emission in each channel to the one derived from the \ns spectral fit, which we call the ``AIA synthetic'' emission. The resulting plot for the emission during the pre-flare and microflare times are shown in Figure~\ref{fig:synobs}. As expected, the hotter emission observed by \ns is only contributing a tiny fraction to the observed emission in most of the \sa channels. The only channels in which the majority of the observed emission is coming from the temperatures \ns observed are, as expected, 94\AA~and \ion{Fe}{18} during the microflare time. This helps confirm why the microflaring loop is only clearly visible in those \sa channels, as there appears to be no significant change in the amount of material at cooler temperatures.

\subsection{Comparison of \ns and {\it GOES}/XRS}

Using the thermal parameters found from fitting the \ns spectra we can estimate the {\it GOES}/XRS flux that should have been produced. For the emission from the whole region during the pre-flare time we estimate the {\it GOES}/XRS flux using the standard routine \textit{goes\_flux49.pro} as $2.7\times10^{-9}$ Wm$^{-2}$. The observed {\it GOES}/XRS flux from the full-disk over this time was actually $6.6\times10^{-8}$ Wm$^{-2}$, about a factor of 25 higher. Similarly, using the \ns temperature and emission found for the microflare excess we obtain a {\it GOES}/XRS flux of $2.0\times10^{-10}$ Wm$^{-2}$, equivalent to 0.02A-class. The observed flux was $1.0\times10^{-8}$ Wm$^{-2}$, about 50 times higher. It could be that there was emission coming from elsewhere on the disk, however close examination of both {\it GOES}/SXI (as shown in Figure~\ref{fig:img_fd}) and \sa \ion{Fe}{18} full disk images show that the \ns region was the main and brightest one on the disk and certainly cannot explain such large discrepancies. The higher flux observed by {\it GOES}/XRS might be due to the presence of emission from lower energies than \ns can detect. However such material would have to be at temperatures just below the ones found with \ns otherwise there would be a clear excess in more \sa channels, not just those sensitive to the hottest material (i.e. 94\AA~and \ion{Fe}{18}).

Although there is a substantial difference between the calculated and observed fluxes it should be noted that {\it GOES}/XRS is poorly calibrated at these low flux levels, as it is designed to monitor large flares. This is highlighted in the recent comparison of {\it GOES}/XRS emission with the softer X-ray spectrometer {\it MinXSS-1} \citep{2016JSpRo..53..328M}. The {\it MinXSS-1} spectrum gives a more robust irradiance measure compared to the broader channel used by {\it GOES}/XRS and showed deviations below fluxes of $10^{-6}$ Wm$^{-2}$, which became even more substantial once below $10^{-7}$ Wm$^{-2}$ \citep{2017ApJ...835..122W}. {\it MinXSS-1} was operational when these \ns observations were made, providing spectra integrated over the full-disk. Unfortunately no event was discernible above the pre-flare level, which may have been due to it operating in a ``non-fine pointing'' mode during this time range \citep[see][]{2019moore}. What {\it MinXSS-1} did observe was consistent with a slightly lower temperature and higher emission measure than the pre-flare one found with \ns, which could help explain the \ns to {\it GOES}/XRS discrepancy during this pre-flare time.

\section{Discussions \& Conclusions}

In this paper, we presented the smallest microflare seen yet with \ns, about an order of magnitude weaker than those previously observed with \ns \citep{2017ApJ...844..132W,2017ApJ...845..122G} and well beyond the microflares observed with {\it RHESSI} \citep{2008ApJ...677..704H}. This event is similar in thermal energy to quiet Sun flares seen with \ns \citep{2018ApJ...856L..32K}, however the microflare presented in this paper demonstrates higher temperature emission and is from an active region. In this microflare we saw emission at about 5MK, which gave an instantaneous thermal energy of around $10^{26}$ erg. It is remarkable that even in this small X-ray microflare we were still able to see corresponding emission in UV, allowing us to study both the coronal and upper chromospheric/transition region response. The small loop seen with \ir in UV and \sa in EUV by itself was unexciting, but this changes with the unexpected addition of emission seen at higher energies with \ns. In this microflare no higher temperature (closer to 10MK) or non-thermal emission was observed but that could be due to limited effective area from only one of the two telescopes observing the flare and as well as the short exposure time. Only about 3s was achieved over an on-time of 60s, due to emission elsewhere on the solar disk. Further \ns observations with higher livetimes will be better able to address the presence of non-thermal emission and/or higher temperatures in events such as this. Observations of small flares have the inherent problem that these are short duration events, so long exposures are not possible and require instruments with higher sensitivity from larger detector effective area. 

It is surprising that this microflare is only seen at the hotter coronal temperatures and lower chromospheric/transition region ones, but there is no increase in emission from material in the few MK range. The \sa channels sensitive to these temperatures show consistent emission during the pre- and microflare times, but no clear excess. It could have been that there was more background material in this temperature range so the small increase due to the microflare was hidden, rendering it effectively invisible. Or it may have been that hotter material seen by \ns and \sa \ion{Fe}{18} cooled too rapidly to be seen, or that the ionisation timescale was longer than the cooling timescale. This event did not present the moss brightenings reported in previous \ir small flare work \citep{2014Sci...346B.315T}, so it could be that this event is even weaker, with faster rastering required to catch velocities clearly associated with the microflare, or possibly a different type of event.

Although the microflare is seen as a brightening in {\it GOES}/XRS, it is difficult to trust the observed flux given that this is at the limit of the instrument's sensitivity and prone to substantial uncertainties in the calibration (in terms of the spectral distribution of these small events relative to the instruments response function). But again it should be noted that {\it GOES}/XRS was not designed to be useful for these small fluxes. Future observations with \ns that overlap with other softer X-ray spectrometers, such as {\it MinXSS-2} \citep{2018SoPh..293...21M} or {\it MaGIXS} \citep{2018SPIE10699E..27K}, might help to resolve the true multi-thermal emission of these small microflares over this energy range.

The \ns observations of this small microflare have shown that even fairly ordinary features seen in UV and EUV can have a higher energy X-ray component. This shows that there is substantial potential for studying weaker solar activity at higher energy X-rays, either occasionally with \ns or with an optimised solar spacecraft such as the proposed {\it FOXSI} \citep{2017arXiv170100792C}.

\acknowledgments
This paper made use of data from the NuSTAR mission, a project led by the California Institute of Technology, managed by the Jet Propulsion Laboratory, funded by the National Aeronautics and Space Administration. We thank the NuSTAR Operations, Software and Calibration teams for support with the execution and analysis of these observations. This research made use of the NuSTAR Data Analysis Software (NUSTARDAS) jointly developed by the ASI Science Data Center (ASDC, Italy), and the California Institute of Technology (USA). 

IRIS is a NASA small explorer mission developed and operated by LMSAL with mission operations executed at NASA Ames Research center and major contributions to downlink communications funded by ESA and the Norwegian Space Centre.

IGH is supported by a Royal Society University Fellowship. The authors thank the International Space Science Institute (ISSI) for support for P. Testa's team ``New Diagnostics of Particle Acceleration in Solar Coronal Nanoflares from Chromospheric Observations and Modeling'', where this work benefited from productive discussions.
\facilities{NuSTAR, IRIS, SDO/AIA, GOES}

%\bibliographystyle{aasjournal}
%\bibliography{refs.bib}

\end{document}